\documentclass[12pt]{iopart}
\usepackage{graphicx}
\begin{document}

\title{A new first principles approach to calculate phonon spectra of disordered alloys}

\author{Oscar Gr\aa n\"as$^1$, Biswanath Dutta$^2$, Subhradip Ghosh$^2$ and 
Biplab Sanyal$^1$}

\address{$^1$ Department of Physics and Astronomy, Uppsala University, Box-516, SE-75120 Uppsala, Sweden}
\address{$^2$ Department of Physics, Indian Institute of Technology Guwahati,
Guwahati-781039,India}
\ead{biplab.sanyal@physics.uu.se}

\date{\today}

\begin{abstract}
The lattice dynamics in substitutional disordered alloys with 
constituents having large size differences is driven by strong
disorder in masses, inter-atomic force constants and local environments.
In this letter, a new first-principles approach based on special quasi random structures
 and itinerant coherent potential approximation to compute the phonon spectra
of such alloys is proposed and applied to Ni$_{0.5}$Pt$_{0.5}$ alloy. 
The agreement between our results with
the experiments is found to be much better than for previous models of disorder 
due to an accurate treatment of the 
interplay of inter-atomic forces among various pairs of chemical species.
This new formalism serves as a potential
solution to the longstanding problem of a proper microscopic understanding of
lattice dynamical behavior of disordered alloys.
\end{abstract}

\submitto{\JPCM}
\maketitle

The energy dispersion of lattice waves in ordered intermetallics
and disordered alloys has recently gained significant importance due to
 its role in stabilizing a particular structural phase by contributing
 towards the system's entropy and in
pinpointing the fundamental mechanism behind the behavior of materials,
which are important in technological applications. Examples include the
connection between martensitic transformations in shape-memory alloys and the
anomalous behavior of their phonon spectra \cite{sma}, anomalously low thermal-expansion coefficients in Fe-based Invar alloys and the unusual
behavior of phonon frequencies under external pressure \cite{invar}, and 
the dramatic
alteration of relative stabilities of different precipitates in Al-Cu system
due to vibrational entropy \cite{pt}. 

Theoretical investigation on lattice dynamics in disordered metallic alloys is
one of the challenging areas in the state of the art materials science
research. Most of the theoretical studies have been done 
on the alloy compounds although a wealth of neutron-scattering data 
is available for various disordered alloys with arbitrary compositions. 
The dearth of theoretical works is due to the lack of a suitable self-consistent, analytical methodology
to carry out the averaging over the random configurations, taking into
account the disorder due to the fluctuations in mass, force constant
and local environment around an atom. The most widely used method 
 is the single-site coherent potential 
approximation (SCPA) \cite{cpa}, which being a single-site theory, is 
unable to handle the non-local fluctuations. The recently developed itinerant
coherent potential approximation (ICPA) \cite{icpa1}, is a 
successful generalization of
the SCPA in handling disordered alloys with large mass as well as 
force constant disorder.
While the ICPA stands as a self-consistent, analytic, site-translational
invariant tool for computing the phonon spectra in disordered alloys,
a missing key component in establishing it as an accurate and reliable
formalism, is its integration with a tractable structural model for
positional disorder which can closely mimic the fluctuations in the 
inter-atomic force constants in a random alloy. 
Moreover, due to the presence of short-range order, the phonon spectra of such
systems would be affected due to the dominant vibrations of
unlike species pairs.
Initially, attempts were
made in the form of first-principles computation of inter-atomic 
force constants from selected ordered structures and use them 
in a random alloy environment \cite{icpa2,icpa3}. 
This is not a proper remedy as the force constants are not directly transferable across the 
environments \cite{Sluiter,vandewalle}.

In this letter, we present a new first-principles based formalism to
compute the lattice dynamics of substitutional disordered alloys which
incorporates the effects of disorder in mass, force constant and
environment. We demonstrate the formalism by computing the phonon dispersion
spectra of Ni$_{0.5}$Pt$_{0.5}$ alloy. This alloy is chosen as its constituents have
large mass differences(Pt being 3 times heavier than Ni), large size
differences ($\sim$ 11$\%$) and large force constant
differences (Pt-Pt force constants being $\sim$ 55 $\%$ larger than the
Ni-Ni ones) and thus significant effects of all three kinds of disorder
on the lattice dynamics can be expected. Also, the neutron-scattering
experiments \cite{ex} show anomalous features in the 
dispersion curves in the form
of resonance modes and splitting of dispersion branches. In previous works,
neither SCPA \cite{ex} nor ICPA \cite{icpa1} with empirical set of 
force constants could have
good quantitative agreement with the experiments, signifying that the
inter-atomic interactions have not been modeled accurately.

Our formalism has {\bf three} important steps: first we generate a structural model of
substitutional disorder by special quasirandom structure (SQS) method, 
then the averaging of the force constant
tensor is done to recover the original symmetry of the solid solution followed by
the ICPA for performing the averaging over configurations.

(i) {\it Structural model}: 
First principles supercell calculations of disordered alloys involve the 
direct averaging of the properties obtained from a number of disordered 
configurations of the alloy. Current computational capabilities limit the size of the supercell 
necessary to describe each
configuration, as well as the number of configurations sampled.
Zunger {\it et. al.} developed a computationally
tractable approach to model the disorder through the introduction of 
SQS method \cite{sqs}, an $N$-atom per cell periodic
structure designed so that their distinct correlation functions 
$\Pi_{k,m}$
best match the ensemble-averaged correlation-functions $\ll \Pi_{k,m} \gg$
of the random alloy. Here $(k,m)$ corresponds to the figure defined by
the number of $k$ of atoms located on its vertices ($k$=2, 3, 4.... are pairs,
triangles, tetrahedra etc.) with $m$ being the order of neighbor distances
separating them ($m$=1, 2....are first,second neighbors etc.). 
As the properties like the 
equilibrium volume, local density of states and bond lengths around an
atom are influenced by the local environment, SQS forms an
adequate approximation. The biggest advantage of the SQS over a
conventional supercell to model positional disorder is that the former
uses the knowledge of the pair-correlation functions, a key property of
the random alloys, to decide the positions of the atoms in the unit cell,
instead of inserting them randomly as is done in the conventional
supercell technique, and thus is guaranteed to provide a better
description of the environments in an actual random alloy.

(ii) {\it Averaging procedure}:  The force constant tensors between a given pair of atoms 
calculated using the SQS will have
nontrivial off-diagonal elements due to low symmetry. 
Moreover, due to the atomic relaxations in a local environment,
a distribution of bond distances for a pair of species gives rise to a distribution in their force constants.
To extract the force constant tensors having the symmetry of the
underlying crystal structure of the disordered alloy, we average the 
calculated SQS force-constant matrices for
a particular set of displacements of the atoms from their ideal positions
in the lattice, related by symmetry that will transform the force constant
matrices to one displaying the symmetry of the underlying crystal
structure. These averaged force-constant tensors are then
used as the inputs to the ICPA for the calculation of the configuration-averaged
quantities. Below, we demonstrate the idea for a FCC lattice.

In a FCC lattice, the distances between a given atom and its 12 nearest 
neighbors are specified by the vectors $(\pm\frac{1}{2},\pm\frac{1}{2},0)a$,
$(\pm\frac{1}{2},0,\pm\frac{1}{2})a$ and $(0,\pm\frac{1}{2},\pm\frac{1}{2})a$; 
$a$ being the lattice parameter. The force constant matrix for two atoms 
separated by the vector $(\frac{1}{2},\frac{1}{2},0)a$ is of the form 

 \[ \left( \begin{array}{ccc}
         A& B&0 \\
         B& A& 0\\
         0& 0& C
        \end{array}\right) \]

The other nearest neighbor force constant matrices are of the same form and 
are related by point group operations. However, when the atoms are allowed to
relax, the vector separating a pair of nearest neighbor atoms are modified to 
$(\frac{1}{2} \pm \delta_{1},\frac{1}{2} \pm \delta_{2},\delta_{3})a$ and the 
matrix of force constants corresponding to a given pair of atoms reflects the
loss of symmetry, taking the general form
\begin{equation}
\Phi = \left( \begin{array}{ccc}
         a_{1}& b_{1}&a_{3} \\
         b_{2}& a_{2}& a_{4}\\
         a_{6}& a_{5}& c
        \end{array}\right)
        \label{eq:mat1}
\end{equation}
  
For a particular set of $\delta_{1}$, $\delta_{2}$ and $\delta_{3}$, one needs
to perform the averaging such that we obtain the following force constant 
matrix:

\begin{equation}
\Phi^{'} = \left( \begin{array}{ccc}
         a^{'}& b^{'}&0 \\
         b^{'}& a^{'}& 0\\
         0& 0 & c^{'}
        \end{array}\right)
\end{equation}

To achieve this, one needs to pick up the relevant ones out of the $48$ 
symmetry operations corresponding to the cubic group which transform the force
constant matrices to one displaying the symmetry of the FCC structure. To 
elaborate on this, we provide the following example:

Suppose, two atoms are separated by the separation vector 
$\vec{R}=\{\frac{1}{2}-\delta_{1},-(\frac{1}{2}+\delta_{2}),-\delta_{3}\}a$. 
The force constant matrix obtained from first-principles calculations on the
SQS is given by Eq.~\ref{eq:mat1}. To recover the symmetry of the FCC force 
constants, we need to find out the transformations that transform the 
separation vector $\vec{R}$ to a new separation vector 
$\vec{R}^{'}=\{\frac{1}{2}-\delta_{1},\frac{1}{2}+\delta_{2},-\delta_{3}\}a$ 
and  $\{\frac{1}{2}+\delta_{2},\frac{1}{2}-\delta_{1},-\delta_{3}\}a$. These
transformations, in this particular case, mimic the nearest neighbor separation
$(\frac{1}{2},\frac{1}{2},0)a$ for the unrelaxed FCC lattice. They are:
\begin{equation}
\begin{array}{ccc}
(x,y,z) & \longrightarrow & (y,-x,z) \\
(x,y,z) & \longrightarrow & (x,-y,z) \\
(x,y,z) & \longrightarrow & (y,-x,-z) \\
(x,y,z) & \longrightarrow & (x,-y,-z).
\end{array}
\end{equation}

Corresponding to each of these transformation, there is a transformation matrix 
$U$. We transform the SQS force-constant matrix $\Phi$ to the FCC 
force-constant matrix $\Phi^{\prime}$ by performing the operation $U^{T}\Phi U$
in each of the 4 cases and then adding them. This simple procedure produces 
$a^{'}=2(a_{1}+a_{2})$, $b^{'}=-2(b_{1}+b_{2})$ and $c^{'}=4c$. All other 
off-diagonal elements vanish due to this symmetrization and the FCC symmetry is
recovered. This procedure is repeated for all pairs of force constants and an 
arithmetic average is finally computed.

(iii){\it The ICPA:} The ICPA is a Green's function based formalism
that generalizes the SCPA by considering scattering from more than
one site embedded in an effective medium within which the effect of this
correlated disorder is built in. The medium is constructed in a self-consistent
way so that site translational invariance and analyticity of the
Green's function are ensured. The analyticity of the Green's function is
ensured by using the principles of the traveling cluster approximation
\cite{tca} which showed that once there are non-diagonal terms in the
Green's function, the self-energy must include itineration of the scatterer 
through the sample to preserve analyticity. This means that the physical observables would be
site translationally invariant. This required translational invariance is
ensured by expressing the operators associated with the physical observables
in an extended Hilbert space which accounts for the statistical fluctuations
in site occupancies due to disorder.

We employ first-principles plane wave projector augmented wave method within generalized gradient approximation (GGA) \cite{pbe} as 
implemented in the VASP code \cite{vasp} for the calculation of the Hellman-Feynman forces 
on the atoms in a 64-atom SQS cell (4x4x4 supercell of the primitive fcc unit cell) with the optimized lattice parameter of 3.93 \AA. Our value of the lattice parameter is slightly larger than the experimental value of 3.78 \AA due to the use of GGA.  The SQS generated structure has zero short range order in the first four neighboring shells indicating a homogeneous disorder.
The cut-off energy for the electronic wave functions is 400 eV. 432 k-points were used in the irreducible Brillouin zone for the calculations of the force constant matrix. We have tested the convergence of our results with respect to the  relevant parameters and have concluded that our choice yields quite accurate results.
To obtain the force-constant matrix, first, the equilibrium geometry is obtained by 
relaxing the atomic positions in the SQS cell until the forces converged to 10$^{-4}$ eV/\AA. 
Then each atom in the SQS cell is moved by 0.01 \AA~from the equilibrium position along three cartesian axes and forces on the atoms are calculated. Due to the lack of any symmetry in the SQS cell, 64 different force constant matrices were generated by using the PHON code \cite{phon}. The symmetry averaging procedure is then used to obtain an effective 3x3 force constant matrix to be used in ICPA calculation. In ICPA, the disorder is considered in the nearest-neighbor
shell only as the further neighbor force constants have one order of
magnitude less. The ICPA calculations are done with a $25 \times 25 \times
25$ $k$-mesh and 1000 energy points.

\begin{figure}[tbp]
\begin{center}
\includegraphics[scale=.35]{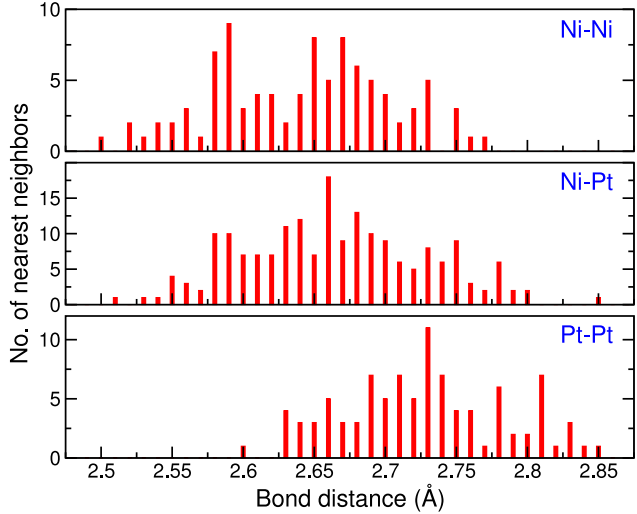}
\caption{\label{fig1}(Color online) Dispersion of bond distances for
three pairs of atoms computed by the SQS.}
\end{center}
\end{figure}
In Fig.~\ref{fig1}, the variations of the number of nearest
neighbors of a given type are plotted as a function of the bond distances. It is observed
that the inter-atomic bond distances depend sensitively on the number of
unlike atoms, which clearly proves the dependence of the bond distances on local
environments. As a result of this, the inter-atomic force constants too
undergo variations. The calculated average Ni-Ni ($d_{Ni-Ni}$), Ni-Pt ($d_{Ni-Pt}$) 
and Pt-Pt ($d_{Pt-Pt}$) bond distances
are 2.64~\AA, 2.66~\AA~and 2.73~\AA~respectively, indicating a
significant dispersion among the three pairs of bonds. A comparison to the unrelaxed
alloy bond distances reveal that the relaxed $d_{Ni-Pt}$ remains the same, 
$d_{Ni-Ni}$ is less than $1\%$ smaller and $d_{Pt-Pt}$ is 
$2.6\%$ larger. The force constants for three different models of disorder, viz., SQS-averaged, 
empirical \cite{icpa1} and SCPA are presented in Table 1.
In Ref.~\cite{icpa1}, it was argued that the Ni-Ni bonds
in the alloy are softer compared to those in pure Ni, the Pt-Pt bonds
are stiffer compared to those in pure Pt and Ni-Pt bonds are even softer
than the Ni-Ni ones because the Ni-Pt bond distances were thought to
be the largest. A comparison between the force constants obtained 
by the SQS-averaged and by this empirical scheme
shows that the Ni-Ni force constants computed by the SQS-averaging scheme are
$50\%$ softer, the Ni-Pt and the Pt-Pt force constants are
$25\%$ and $23\%$ harder on an average. More importantly, the Ni-Pt
force constants computed by the SQS-averaging scheme are harder than the
Ni-Ni ones, a result in contradiction with the empirical scheme. This difference occurs due to the proper
inclusion of environmental disorder in our case. $d_{Ni-Ni}$ in the
alloy as computed by the SQS are about $6\%$ larger than that in pure Ni
and thus these bonds suffer most severe dilution in strength, $d_{Pt-Pt}$ reduces
 by about $1\%$ compared to that in pure Pt and thus
they are harder by about only $15\%$ compared to those in pure Pt.
Inspite of nearly same bond distances, the Ni-Pt
bonds computed by the SQS are stiffer than the Ni-Ni bonds, due to the
following reasons: in case of the Ni-Pt pairs, the Ni atoms find much
larger Pt atoms as their nearest neighbors, roughly with the same
available space as they get in case of Ni-Ni pairs. As a result, the
smaller Ni atoms try to accommodate bigger Pt atoms within the same volume
as that available for Ni-Ni pairs, resulting in a hardening of Ni-Pt
nearest neighbor interactions compared to the Ni-Ni ones.

\begin{table}
\caption{\label{table1}Real-space nearest neighbor force constants 
$\Phi^{\alpha \beta}_{s s^{\prime}}$ (in dyn cm$^{-1}$) for Ni$_{50}$Pt$_{50}$
obtained by averaging the SQS force constants, the force constant matrix 
$\Phi^{\alpha \beta}$ used for SCPA calculations and the empirical force 
constants \cite{icpa1}. The last column indicates the cartesian components of 
$\Phi^{\alpha \beta}$.}
\begin{indented}
\item[]\begin{tabular}{@{}lllll}
\br
 Pair &  SQS & SCPA & Empirical & $\alpha \beta$ \\
\mr
Ni-Ni  &   -8231 & -19365 & -15587 & $xx$\\
Ni-Pt  &  -17868 & " & -13855 & " \\
Pt-Pt  &  -33494 & " & -28993 & " \\
\mr
Ni-Ni  &     525 & 3255 & 436 & $zz$\\
Ni-Pt  &    2820 & " & 348 & " \\
Pt-Pt  &    6854 & " & 7040 & " \\
\mr
Ni-Ni  &  -9580  & -22679 & -19100 & $xy$\\
Ni-Pt  &  -20740 & " & -15280 & " \\
Pt-Pt  &  -39655 & " & -30317 & " \\
\br
\end{tabular}
\end{indented}
\end{table}

\begin{figure}[tbp]
\begin{center}
\includegraphics[scale=.35]{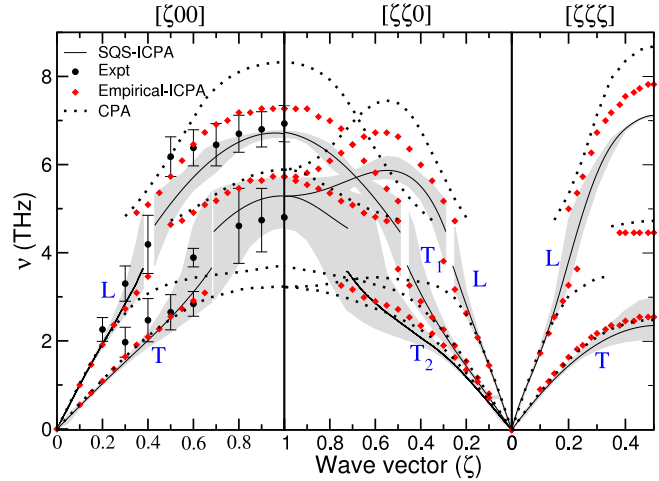}
\caption{\label{fig2}(Color online) Phonon dispersion in Ni$_{0.5}$Pt$_{0.5}$
alloy computed by SQS-ICPA (solid line), SCPA(dotted line) and
empirical-ICPA \cite{icpa1}(red diamonds). The circles indicate the
experimental results. The shaded regions indicate the disorder-induced
widths calculated by the SQS-ICPA.}
\end{center}
\end{figure}

In Fig.~\ref{fig2}, we present the phonon dispersion curves of
Ni$_{0.5}$Pt$_{0.5}$ alloy computed using force constants
obtained by three models
of disorder. 
The configuration averaging in case of the
SQS and the empirical scheme are done by the 
ICPA. SQS-ICPA shows the best
agreement with the experiments as compared to the SCPA and the
empirical-ICPA \cite{icpa1}.The disorder-induced widths
computed by the SQS-ICPA method (the shaded region in Fig.~\ref{fig2}) also
agree reasonably well with the experiments. The comparisons 
with the experiments were possible only for the $[\zeta 0 0]$ 
direction 
because experimental results were not available for other directions.
A closer look at Fig.~\ref{fig2} reveals that for the low-frequency branches,
the empirical-ICPA and the SQS-ICPA
methods are in close agreement while there are significant differences
for the high frequency
branches. The mass-disorder treated in SCPA, on the other hand, fails to 
reproduce the experimental features of the dispersion relations both
qualitatively and quantitatively. The frequencies of the high frequency
branches (both transverse and longitudinal) computed by the SCPA are
severely overestimated while the lower frequency branches extend all
the way to the zone boundary, thus displaying a split-band behavior in
the phonon dispersions which is not observed experimentally. 

All these observations can be understood in terms of the inter-atomic force
constants (shown in Table I) which are influenced by the fluctuations in the local
environments. In the SCPA, the fluctuations in the force constants
are completely neglected and hence Ni-Ni, Ni-Pt and Pt-Pt all have the same
force constants. In earlier studies \cite{ex,icpa1}, the force constants
used in the SCPA were those of pure Ni ones obtained experimentally
\cite{niexp}. In this study we have used a more realistic set of
force constants for the SCPA calculations which are $\Phi^{\alpha \beta}=
x^{2}\Phi^{\alpha \beta}_{Ni-Ni}+\left(1-x \right)^{2}\Phi^{\alpha \beta}_{Pt-Pt}+2x\left(1-x \right)\Phi^{\alpha \beta}_{Ni-Pt}$ where $\alpha, \beta$ are
the cartesian directions and $x$ is the concentration of Ni.  $\Phi^{\alpha \beta}_{s s^{\prime}}$ are the SQS-averaged force constants where $s, s^{\prime}$ denote atomic species. Such a 
choice of force constants has been used to incorporate, in an average way, the effects of
alloy environment. The results, nevertheless, suggest 
that unless the fluctuations in the force constants are incoporated, the results do not even agree
qualitatively with the measurements. The frequencies of the upper (lower) branches
of longitudinal and transverse modes computed by SCPA are too high (low) as
compared to the experiments. Ni-Ni
vibrations dominate the higher frequency branches and the average 
force constants being too high pushes the frequencies away from the
ones measured by neutron-scattering while the lower frequency branches being
solely due to the Pt-Pt vibrations have lower frequencies due to the underestimation of the Pt-Pt
force constants. 
The empirical-ICPA results on the other hand agree qualitatively with the
experiments because there is no split-band like behavior as was seen in the
SCPA. This is due to the consideration of the Ni-Pt correlated 
vibrations which renormalize the spectral weights associated with the
contributions from Ni-Ni and Pt-Pt pairs \cite{icpa1}.
Moreover, the splitting of the vibrational branches found experimentally around 
$\zeta=0.55$ as a signature of the existence of resonance mode, is reproduced and
the frequencies of the high-frequency branches are better as
compared to the SCPA. The Ni-Ni(Pt-Pt) force constants shown in Table I in the
empirical model are softer(harder) as compared to the SCPA ones explaining
the reason for better agreement of the phonon frequencies computed by
the empirical-ICPA model with the experimental results. 
However, the overestimation(underestimation) of the Ni-Ni(Pt-Pt) 
interactions and an incorrect qualitative estimation of the Ni-Pt
interactions as compared to the Ni-Ni ones gives rise to significant discrepancies 
for the high-frequency longitudinal and optical branches. 
With SQS-ICPA, one can have a much better
quantitative agreement between theory and experiments as seen in 
Fig.~\ref{fig2}.
The high-frequency branches for
both longitudinal and transverse vibrations computed by the SQS-ICPA
method agree substantially with the experimental results. The normal mode 
frequencies for these branches are dominated by
the vibrations of the Ni pairs and thus a softening of the Ni-Ni bonds
as computed by the SQS pushes the frequencies downwards compared to the 
empirical model making a better agreement with the experiments. Similarly,
the relaxations of the Pt atoms result in the stiffening of the Pt-Pt bonds, thus pushing the
frequencies slightly upwards. However, the high-frequency transverse branch 
computed by the SQS-ICPA model is still overestimated. 
In the neutron-scattering measurements
\cite{ex}, there were some ambiguities in determining the peak positions of
the line shapes for the high-frequency transverse modes and thus the 
experimental results for this branch had larger uncertainties. Given this
fact, the agreement between the theory and the experiment can be considered
to be fairly good. 

In conclusion, we have developed a reliable first-principles based approach for the
calculation of phonon spectra in substitutional disordered alloys to
treat mass, force constant and environmental disorder on equal footing. 
We demonstrate in case of Ni$_{0.5}$Pt$_{0.5}$ alloy,
the importance of an accurate structural model of disorder taking into account the role 
of fluctuations in the local environment through atomic relaxations in interpreting the
microscopic features of the lattice dynamics for this class of complex
alloys. The accurate modeling of the environmental disorder made
possible by the SQS paves the way for a reliable description of phonon
spectra in alloys with short-range order where the force-constants
between a pair of species is dominated by a particular configuration of
the nearest neighbor environment around an atom. Our future aim is to calculate the thermodynamic quantities \cite{grabowski} to compare with the experiments. Finally, we conclude that a combination of reliable force constants obtained from ab-initio and
ICPA as a self-consistent analytic method for configuration-averaging enables us to 
solve the longstanding problem of theoretical computation
of lattice dynamics in disordered alloys.

\ack

BS acknowledges G\"{o}ran Gustafssons Stiftelse and Swedish Research Council 
for financial support and SNIC-UPPMAX for granting computer time. BD 
acknowledges CSIR, India for financial support under the Grant-F. No. 09/731 
(0049)/2007-EMR-I. Also, we acknowledge Prof. P. L. Leath, Prof. M. H. Cohen, 
Dr. J. B. Neaton and Dr. A. H. Antons for useful discussions.


\section*{References}
\begin{thebibliography}{99}
\bibitem{sma} Bungaro C and Rabe K M 2003 {\it Phys. Rev. B} {\bf 68} 134104.
\bibitem{invar} Noda Y and Endoh Y 1988 {\it J. Phys. Soc. Jpn.} {\bf 57} 4225.
\bibitem{pt} Wolverton C and Ozolins V 2001 {\it Phys. Rev. Lett.} {\bf 86} 
5518.
\bibitem{cpa} Taylor D W 1967 {\it Phys. Rev.} {\bf{156}} 1017.
\bibitem{icpa1} Ghosh S, Leath P L and Cohen M H 2002 {\it Phys. Rev. B} {\bf 
66} 214206.
\bibitem{icpa2} Ghosh S, Neaton J B, Antons A H and Cohen M H 2004 {\it Phys. 
Rev. B} {\bf 70} 024206.
\bibitem{icpa3} Alam A, Ghosh S and Mookerjee A 2007{\it Phys. Rev. B} {\bf 75}
134202.
\bibitem{Sluiter} Sluiter M H F, Weinert M and Kawazoe Y 1999 
{\it Phys. Rev. B} {\bf 59} 4100.
\bibitem{vandewalle} van de Walle A and Ceder G 2002 {\it Rev. Mod. Phys.} {\bf
74} 11. 
\bibitem{ex} Tsunoda Y, Kunitomi N, Wakabayashi N, Nicklow R M and Smith H G 
1979 {\it Phys. Rev. B} {\bf 19} 2876. 
\bibitem{sqs} Zunger A, Wei S -H, Ferreira  L G and Bernard J E 1990 {\it Phys. Rev. Lett.} {\bf 65} 353; Lu Z W, Wei S -H and Zunger A 1991 {\it Phys. Rev. B} {\bf 44} 
10470.
\bibitem{tca} Mills R and Ratanavararaksha P 1978 {\it Phys. Rev. B} {\bf 18} 
5291.
\bibitem{pbe} Perdew J P, Burke K and Ernzerhof M 1996 {\it Phys. Rev. Lett.} {\bf 77} 3865.
\bibitem{vasp} Kresse G and Hafner J 1993 {\it Phys. Rev. B} {\bf 47} RC558; 
Kresse G and Furthm\"{u}ller J 1996 {\it Phys. Rev. B} {\bf 54} 11169.
\bibitem{phon} Alf\'{e} D 2009 {\it Comp. Phys. Comm.} {\bf 180} 2622.
\bibitem{niexp} Dutton D H 1972 {\it Can. J. Phys.} {\bf 50} 2915.
\bibitem{grabowski} Grabowski B, Hickel T and Neugebauer J 2007 {\it Phys. Rev. B} {\bf 76} 024309. 
\end{thebibliography}
\end{document}